# Comparative study of Authorship Identification Techniques for Cyber Forensics Analysis

Smita Nirkhi
Department of Computer Science & Engg
G.H.Raisoni College of Engineering
Nagpur, India

Dr.R.V.Dharaskar
Director
MPGI
Nanded, India

*Abstract*—**Authorship Identification techniques are used to identify the most appropriate author from group of potential suspects of online messages and find evidences to support the conclusion. Cybercriminals make misuse of online communication for sending blackmail or a spam email and then attempt to hide their true identities to void detection.Authorship Identification of online messages is the contemporary research issue for identity tracing in cyber forensics. This is highly interdisciplinary area as it takes advantage of machine learning, information retrieval, and natural language processing. In this paper, a study of recent techniques and automated approaches to attributing authorship of online messages is presented. The focus of this review study is to summarize all existing authorship identification techniques used in literature to identify authors of online messages. Also it discusses evaluation criteria and parameters for authorship attribution studies and list open questions that will attract future work in this area.**

## I. INTRODUCTION

Cyber crime is also known as computer crime, the use of a computer to further illegal ends, such as committing fraud, trafficking in child pornography and intellectual property, stealing identities, or violating privacy.

Cybercrime, especially through the Internet, has grown in importance as the computer has become central to commerce, entertainment, and government. Senders can hide their identities by forging sender's address; Routed through an anonymous server and by using multiple usernames to distribute online messages via different anonymous channel.

Author Identification study is useful to identify the most plausible authors and to find evidences to support the conclusion.

Authorship analysis problem is categorized as [13]

*1) Authorship identification (authorship attribution): It determines the likelihood of a piece of writing to be produced by a particular author by examining other writings by that author.*

*2) Authorship characterization: It summarizes the characteristics of an author and generates the author profile based on his/her writings like Gender, educational, cultural background, and writing style*

*3) Similarity detection: It compares multiple pieces of writing and determines whether they were produced by a single author without actually identifying the author like Plagiarism detection. To extract unique writing style from the number of online messages various features need to be considered are Lexical features, content-free features, Syntactic features ,Structure features ,Content-specific features*

Although authorship attribution problem has been studied in the history but in the last few decades, authorship attribution of online messages has become a forthcoming research area as it is confluence of various research areas like machine learning, information Retrieval and Natural Language Processing. Initially this problem started as the most basic problem of author identification of anonymous texts (taken from Bacon, Marlowe and Shakespeare) [1], now has been grown for forensic analysis, electronic commerce etc. This extended version of author attribution problem has been defined as *needle-in-a-haystack* problem in [2]

When an author writes they use certain words unconsciously and we should able to find some underlying pattern for an authors style. The fundamental assumption of authorship attribution is that each author has habit of using specific words that make their writing unique Extraction of features from text that distinguish one author from another includes use of some statistical or machine learning techniques.

Rest of the Paper is organized as follows. Section 2 Reviews existing techniques used for Authorship Analysis along with their classification. Section 3 explains basic procedure for authorship analysis. Section 4 summarizes Comparisons of various techniques since year 2006 till 2012.Section 5 Reviews performance evaluation parameters required for Authorship Analysis Techniques followed by section 6 which is conclusion.

## II. STATE OF THE ART OF CURRENT TECHNIQUES

This section gives fundamental idea on existing Authorship Attribution Techniques followed by their comparison in next section. In literature, this problem was solved using statistical Analysis and Machine learning techniques. These are mainly categorized as shown in Figure 1.

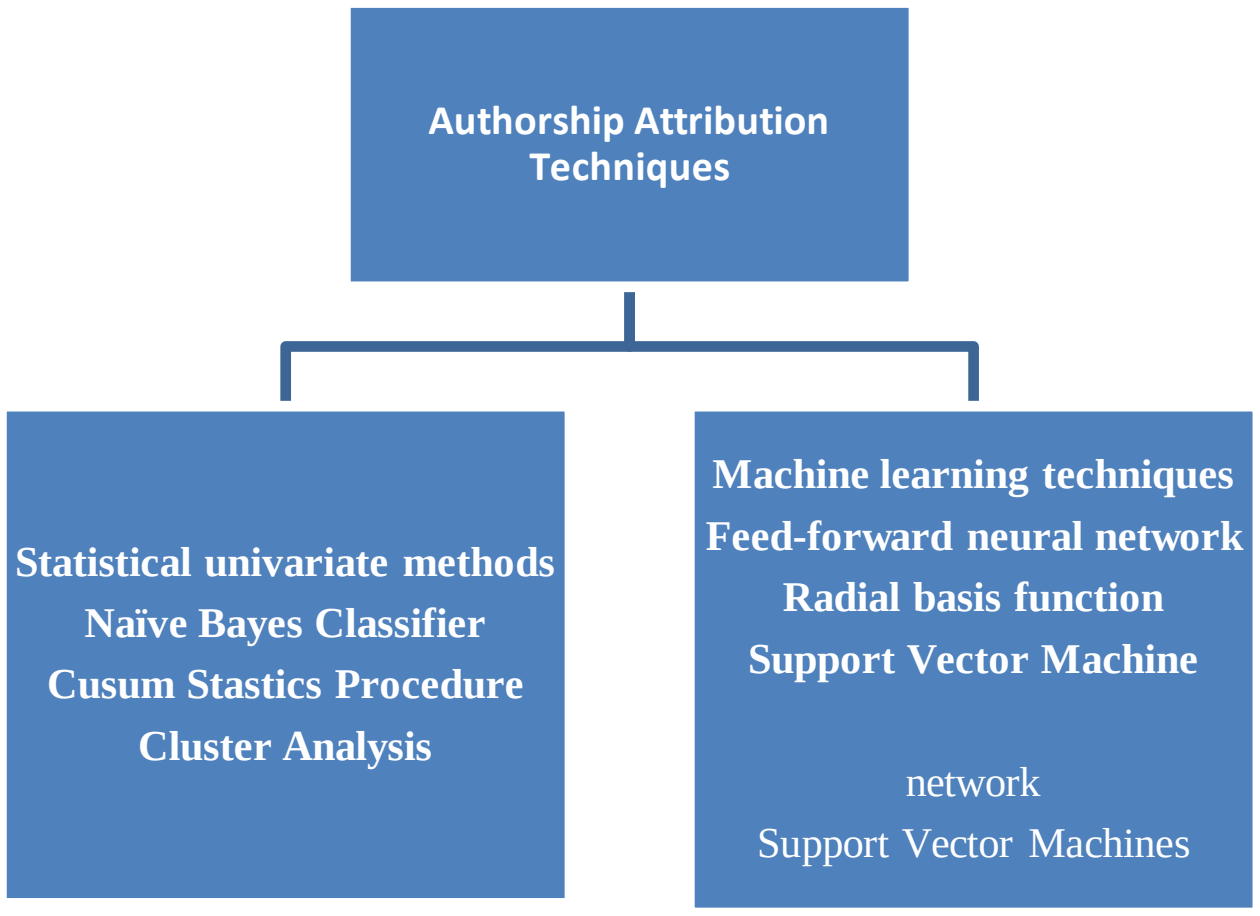

Fig. 1. Authorship Attribution Techniques

STATISTICAL UNIVARIATE METHODS

A) *Naive Bayes classifier*: In this Classifier Learning and classification methods based on probability theory. In Literature it is found that Bayes theorem plays a critical role in probabilistic learning and classification. It uses prior probability of each category given no information about an item.

B) *B.CUSUM statistics procedure*: In stastical analysis the cusum called cumulative sum control chart, the CUSUM is a sequential Analysis technique used for onitoring change detection. As its name implies, CUSUM involves the calculation of a cumulative sum.

C) *Cluster Analysis:* Cluster analysis is an exploratory data analysis tool for solving classification problems. Its purpose is to sort cases (people, things, events, etc) into groups, or clusters, so that the degree of association is strong between members of the same cluster and weak between members of different clusters.

## III. MACHINE LEARNING TECHNIQUES

### A. *Feed-forward neural network :*

A feed forward neural network is an artificial neural network where connections between the units do *not* form a directed cycle. This is different from networks. The feed forward neural network was the first and arguably simplest type of artificial neural network devised. In this network, the information moves in only one direction, forward, from the input nodes, through the hidden nodes (if any) and to the output nodes. There are no cycles or loops in the network.

### B. *Radial basis function network:*

A radial basis function network is an artificial neural network that uses radial basis functions as activation functions. The output of the network is a linear combination of radial basis functions of the inputs and neuron parameters. Radial basis function networks are used for function approximation, time series prediction, and system control.

### C. *Support Vector Machines:*

In machine learning, support vector machines (SVMs, also support vector networks are supervised learning models with associated learning algorithms that analyze data and recognize patterns, used for classification and regression analysis. The basic SVM takes a set of input data and predicts, for each given input, which of two possible classes forms the output, making it a non-probabilistic binary linear classifier.

## IV. CLASSIC PROCEDURE FOR AUTHORSHIP IDENTIFICATION

Figure 2 shows classic approach to model authorship identification problem.

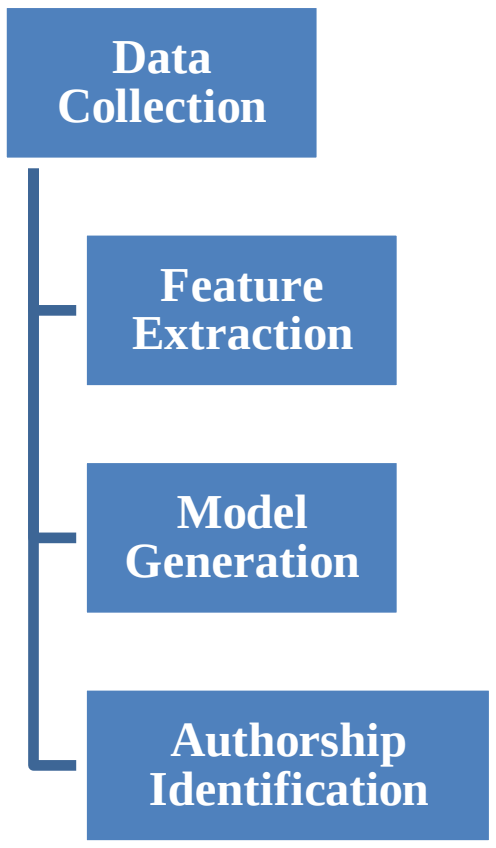

Fig. 2. Typical Procedure for Authorship Identification

Step1: Data collection:-Collect online messages written by potential authors from online communication.

Step2: Feature Extraction:-After extraction, each unstructured text is represented as a vector of writing-style features

Step3: Model Generation:-Dataset should be divided into training and testing set. Classification techniques should be applied. An iterative training and testing process may be needed

Step4: Author Identification:-Developed model can be used to predict the authorship of unknown online messages

## V. COMPARISON OF VARIOUS TECHNIQUES

This section compares the various techniques used for authorship identification research forum since 2006 to 2012.History of studies on authorship attribution problems presented in tabular format and year wise. For each method, we identify the corpus on which methods were tested, the feature types used and the categorization method used, size of Training set. Table 1 represented the comparative study of all authorship techniques.[5][6][7][8][9][10].

| YEAR/AUTHORS | FEATURES | TECHNI QUES | CORPUS | NUMBER OF AUTHORS | TRAINING SET |
|---|---|---|---|---|---|
| **(2006)**<br>**Rong Zheng,**<br>**Jiexun Li, Hsinchun Chen, Zan Huang** | **Lexical,**<br>**syntactic,**<br>**structural,**<br>**content Specific** | **SVM** | **English Internet newsgroup messages & Chinese Bulletin Board System (BBS) messages.** | **20** | **48 for English**<br>**37 (Chinese)** |
| **2006**<br>**Ahmed Abbasi and Hsinchun Chen** | **Lexical,**<br>**syntactic,**<br>**structural,**<br>**content Specific** | **PCA** | **USENET forum, Yahoo group forum , website forum for the White Knights** | **10** | **30 msgs per forum** |
| **2007**<br>**cyran** | **Lexical,**<br>**syntactic,** | **ANN** | **Novels of two famous Polish writers, Henryk Sienkiewicz and Bolesław Prus** | **2** | **168** |
| **2007**<br>**Daniel Pavelec, Edson Justino, and Luiz S. Oliveira** | **Linguistic Features** | **SVM** | **Our sources were two dif- ferent Brazilian newspapers, Gazeta do Povo (http://www.gazetadopovo.com.br) and Tribuna do Paran´** | **10** | **150** |
| **2008**<br>**EFSTATHIOS STAMATATOS** | **Stylistic Fearures** | **SVM** | **Corpus Volume 1 (RCV1)**<br>**Arabic Corpus:** | **10** | **1000** |
| **Kim Luyckx and Walter Daelemans** | **Syntactic Features** | **Memory based learning approac** | **Personae corpus** | **145** | **1400 words** |
| **2008**<br>**Chun Wei** | **Email features** | **clustering** | **Email dataset** | **42** | **4200** |
| **2008(Hamilton)** | **Syntactic Features** | **Stylometroy** | | **145** | **2000** |
| **2008**<br>**Farkhund Iqbal, Rachid Hadjidj, Benjamin C.M. Fung, Mourad Debbabi** | **Stylometric Features** | **Frequent Pattern** | **Enron Dataset** | **158** | **200399** |
| **2008(M.Connor)** | **Syntactic** | **Decision Trees/KNN.** | **Emails collected from users** | **12** | **120** |
| **2009**<br>**Rachid Hadjidj, Mourad Debbabi, Hakim Lounis, Farkhund Iqbal,Adam Szporer, Djamel Benredjem** | **Stylometory Features** | **Stastical Analysis, Machine Learning** | **Enron Dataset** | **158** | **200399** |
| **2011**<br>**George K. Mikros1 and Kostas Perifanos** | **Linguistic features** | **Regularized Logistic Regression (RLR) SVM** | **Dataset** | **-** | **-** |
| **2012**<br>**Ludovic Tanguy, Franck Sajous, Basilio Calderone,** | **Linguistic Features** | **Machine Learning Tool** | **Dataset** | **10** | **100 words** |

## VI. CONCLUSION

The complexity level of aforementioned problem is determined by the various parameters like the number of authors and size of training set. This both the parameters play vital role to determine prediction accuracy. Although these parameters are considered critical to the complexity of the problem and therefore the prediction accuracy, there are no studies examining their impact on the authorship-identification performance in a systematic way. The problem of authorship attribution is explored well in the area of literature, newspapers etc but limited work has been done for the authorship identification of online messages like blogs, emails and chat. This comparative study concluded that if number of author's increases and size of training sets decreases then performance degrades. Thus, by considering all these parameters further research direction is to improve prediction accuracy.

## REFERENCES

[1] Estival 2008] [Abbasi et. al. 2008] [Koppel et. al. 2003] [De Vel et. al. 2001].

[2] Li, J., Chen, H., & Huang, Z. "A Framework for Authorship Identification of Online Messages: Writing-Style Features and classification Technique", *Journal of the American Society for Information Science*, *57*(3), 378–393. doi:10.1002/asi,2006.

[3] Abbasi, A., & Chen, H. "Visualizing Authorship for Identification", *English*, 60–71, (2006).

[4] Stańczyk, U., & Cyran, K. A. "Machine learning approach to authorship attribution of literary texts", *Journal of Applied Mathematics*, *1*(4), 151–158, (2007).

[5] Pavelec, D., Justino, E., & Oliveira, L. S. "Author Identification using Stylometric Features",*Inteligencia Artificial*, *11*(36), 59–65. doi:10.4114/ia.v11i36.892, (2007).

[6] Stamatatos, E. "Author identification: Using text sampling to handle the class imbalance problem", *English*, *44*, 790–799. doi:10.1016/j.ipm.2007.05.012, (2008).

[7] Iqbal, F., Hadjidj, R., Fung, B. C. M., & Debbabi, M. "A novel approach of mining write-prints for authorship attribution in e-mail forensics", *Information Systems*, *5*, 42–51. doi:10.1016/j.diin.2008.05.001, (2008).

[8] Iqbal, F., Binsalleeh, H., Fung, B. C. M., & Debbabi, M. "Mining writeprints from anonymous e-mails for forensic investigation", *Digital Investigation*, 1–9. doi:10.1016/j.diin.2010.03.003, (2010).

[9] Mikros, G. K., & Perifanos, K. "Authorship identification in large email collections: Experiments using features that belong to different linguistic levels, (2011).

[10] Tanguy, L., Sajous, F., Calderone, B., & Hathout, N. "Authorship attribution: using rich linguistic features when training data is scarce", (2012).